\documentstyle[12pt]{article}
\title{
       Initial static susceptibilities \protect \\
       of nonuniform and random Ising chains}
\author{
         Oleg Derzhko, Oles' Zaburannyi\\
\small   {\em {Institute for Condensed Matter Physics,}}\\
\small   {\em {1 Svientsitskii St., L'viv-11, 290011, Ukraine,}}\\
	 J. W. Tucker\\
\small   {\em {Department of Physics,
               The University of Sheffield,}}\\
\small   {\em {Sheffield, S3 7RH, United Kingdom}}}

\date{\today}

\begin{document}

\renewcommand\baselinestretch {1.05}
\large\normalsize

\maketitle

\begin{abstract}
Within the conventional framework
of standard linear response theory
we have derived exact results for the initial
static susceptibilities
of nonuniform spin-$\frac{1}{2}$ Ising chains.
The results obtained permit one to study
regularly alternating-bond
and random-bond Ising chains.
The influence of several types of nonuniformity and disorder
on the temperature dependence
of the initial longitudinal and transverse static susceptibilities
is discussed.
\end{abstract}

\vspace{5mm}


\vspace{5mm}


\vspace{1mm}

\noindent
{\bf Postal addresses:}\\

\vspace{1mm}

\noindent
{\em
Dr. Oleg Derzhko (corresponding author)\\
Oles' Zaburannyi\\
Institute for Condensed Matter Physics\\
1 Svientsitskii St., L'viv-11, 290011, Ukraine\\
Tel: (0322) 42 74 39\\
Fax: (0322) 76 19 78\\
E-mail: derzhko@icmp.lviv.ua}\\

\vspace{1mm}

\noindent
{\em
Dr. J. W. Tucker\\
Department of Physics, The University of Sheffield\\
Sheffield, S3 7RH, United Kingdom\\
Tel: +44 114 222 4273\\
Fax: +44 114 272 8079\\
E-mail: j.w.tucker@sheffield.ac.uk}

\clearpage

\renewcommand\baselinestretch {2.00}
\large\normalsize

\section{Introduction}

During the last 40 years considerable effort has been devoted to
the study
of random magnetic systems.
In order to obtain a theoretical understanding of various
phenomena that can occur due to the presence of disorder it is
instructive to examine simple
models that allow one to calculate observable quantities exactly.
Among such models,
perhaps the simplest is the nonuniform spin-$\frac{1}{2}$
Ising chain specified by the Hamiltonian
\begin{eqnarray}
H_0=\sum_nJ_ns_n^xs_{n+1}^x,
\end{eqnarray}
where the site index $n$ of the summation takes values 1 to $N-1$, there
being a total of $N$ lattice sites.
$J_n$ is the exchange coupling between
the spins at sites
$n$ and $n+1$,
and $s^x$ has eigenvalues $\pm 1/2$.
The partition function of this model is readily calculated, and for
open boundary conditions is
\cite{s}
$Z_0\equiv{\mbox{Tr}}\;[\exp (-H_{0}/kT)]
=2\prod_n[2\cosh(J_{n}/4kT)]$, which
immediately yields the entropy
$S=k\left[
\ln 2 +
\sum_n \ln 2\cosh(J_{n}/4kT)
-\sum_{n}(J_{n}/4kT)\tanh (J_{n}/4kT) \right] $,
the specific heat
$C=k\sum_{n}\left[
(J_{n}/4kT)/\cosh (J_{n}/4kT)\right]^2$,
and the correlation functions
\begin{eqnarray}
\langle
s_m^xs_p^x
\rangle_0
=\frac{1}{4}
\left(-\tanh\frac{J_m}{4kT}\right)
\left(-\tanh\frac{J_{m+1}}{4kT}\right)
\ldots
\left(-\tanh\frac{J_{p-1}}{4kT}\right),
\;\;\; m<p.
\end{eqnarray}
Here, the angular brackets $\langle (\ldots )\rangle_0$ denote
$(1/Z_{0}){\mbox {Tr}}\left[\exp (H_{0}/kT)(\ldots)\right]$.
Regarding the exchange couplings in Eq. (1)
as random variables
having a probability distribution $p(J_{1},\ldots , J_{N-1})$,
one is able to study the
random-averaged quantities
$\overline{S}$,
$\overline{C}$,
or
$\overline{\langle s_m^xs_p^x\rangle _0}$,
defined through
$\overline{Q}\equiv \int \begin{rm}d\end{rm}J_{1}
\ldots \begin{rm}d\end{rm}J_{N-1}\;p(J_{1},\ldots ,J_{N-1})Q$.
In the following,
attention will focus on a study of
the initial (zero-field)
static susceptibilities of nonuniform and random Ising chains
described by Eq. (1).
In the presence of a small external field $h_{\alpha}$, $\alpha=x,y,z$,
static susceptibilities may be defined by
$\chi_{\beta\alpha}(h_{\alpha })\equiv
\partial m_{\beta}/\partial h_{\alpha}$,
$m_{\beta}\equiv (1/N)\sum_n
\langle s_n^{\beta}\rangle $
(in units where
the product of
the g-factor and the Bohr magneton is unity).
The angular brackets $\langle (\ldots )\rangle $ denote thermal averages
with respect to the total Hamiltonian $H=H_{0}+H_{\alpha }$, $H_{\alpha }$
being the Zeeman term $-h_{\alpha }\sum _{n}s_{n}^{\alpha }$.
The initial static susceptibilities are the limits
$\chi _{\beta \alpha }(h_{\alpha })]_{h_{\alpha }\rightarrow 0}$,
to be denoted
by $\chi _{\beta \alpha }$.

The static susceptibilities of random Ising chains
have been studied by a number of authors.
The Ising chain with random exchange coupling in a longitudinal field
was considered by Fan and McCoy \cite{fm}.
Assuming that each exchange coupling was an independent random variable
having a probability density function with narrow width,
Fan and McCoy were able to study the influence of the randomness
on the longitudinal static susceptibility.
The ground-state transverse static susceptibility
for the random-bond Ising chain
was examined by Barouch and McCoy \cite{bm}.
Zaitsev studied the transverse static susceptibility for
the random-bond Ising chain for
various types of disorder
\cite{za}.
The thermodynamics and spin correlations of a model of solid solution
based on an Ising chain were considered in Ref. \cite{bvz}.
Some results for the initial longitudinal and transverse
static susceptibilities
of the random Ising chain in a transverse field
were collected in Ref. \cite{bk}.
Thermodynamic and dynamic properties
of a random Ising chain in a transverse field
for arbitrary disorder may also be examined numerically
\cite{dkv}.
Recently, Idogaki, Rikitoku and Tucker \cite{irt}
have calculated exactly
the initial transverse static susceptibility
for the nonuniform model of Eq. (1),
and for the random version of this model
in which two types of exchange couplings
$J_1$ and $J_2$ are randomly distributed throughout the chain.

The object of our paper is two-fold.
First, in Ref. \cite{irt}
the derivation of the initial static transverse susceptibility
was achieved by a somewhat unconventional technique
based on an accurate extraction of the linear field-dependent term
from a Callen-Suzuki identity for the magnetization.
Here, we show how the result may be obtained more naturally
within the framework of conventional linear response theory
in which the susceptibility is related
directly to the double-time correlation functions.
In addition all components of the susceptibility tensor
will be considered (Section 2).
The second objective of this work
is to apply the theory to chains having various types of disorder,
to identify the extent to which qualitative features of the susceptibility
are influenced by the character of the disorder, and
to compare the influence of regular alternation and random disorder.
Unlike, the bimodial distributions of disorder, hitherto studied
\cite{irt},
more general distributions of disorder cannot be treated analytically,
but require numerical computation. The result of such computations
for chains having Gaussian and Lorentzian disorder
are reported for the first time, in Section 3.

\section{Initial static susceptibilities
        \protect\\
        of nonuniform Ising chains}

Our derivation of the
initial static
susceptibilities of a nonuniform Ising chain employs as its starting
point the well-known expression
\begin{eqnarray}
\chi_{\beta\alpha}(h_{\alpha })=
\frac{1}{kT}
\frac{1}{N}
\sum_{n,p}
\left[
\int_0^1\begin{rm}d\end{rm}\tau
\langle s_n^{\beta}
\left(
-\frac{\begin{rm}i\end{rm}\tau}{kT}
\right)
s_p^{\alpha}\rangle
-\langle s_n^{\beta}\rangle
\langle s_p^{\alpha}\rangle
\right],
\nonumber\\
s_n^{\beta}(t)\equiv{\mbox {e}}^{\begin{rm}i\end{rm}tH}
s_n^{\beta}
{\mbox {e}}^{-\begin{rm}i\end{rm}tH},
\end{eqnarray}
relating static susceptibilities to spin correlation functions.
For the initial static susceptibilities ($h_{\alpha }\rightarrow 0$)
it follows, from Eq. (3), that
\begin{eqnarray}
\chi_{xx}=
\frac{1}{kT}
\frac{1}{N}
\sum_{n,p}
\langle s_n^{x}s_p^{x}\rangle_0,
\end{eqnarray}
and
\begin{eqnarray}
\chi_{zz}=
\frac{1}{kT}
\frac{1}{N}
\sum_{n,p}
\int_0^1\begin{rm}d\end{rm}\tau
\langle \hat{s}_n^{z}
\left(
-\frac{\begin{rm}i\end{rm}\tau}{kT}
\right)
s_p^{z}\rangle_0,\nonumber\\
\hat{s}_{n}^{z}(t)
\equiv \begin{rm}e\end{rm}^{\begin{rm}i\end{rm}tH_{0}}
s_{n}^{\beta }\begin{rm}e\end{rm}^{-\begin{rm}i\end{rm}tH_{0}},
\end{eqnarray}
since
$\langle s_n^{\alpha }\rangle_0=0$.

The integral in Eq. (5) may be effected by employing the Van der Waerden
identity $\exp (\lambda s^{x})
=\cosh (\lambda /2)+2\sinh (\lambda /2)s^x$.
Using this identity, one finds that
\begin{eqnarray}
\hat{s}_j^z(t)
=
\cos
\left[
t
\left(
J_{j-1}s_{j-1}^x+J_js_{j+1}^x
\right)
\right]
s_j^z
+
\sin
\left[
t
\left(
J_{j-1}s_{j-1}^x+J_js_{j+1}^x
\right)
\right]
s_j^y,
\end{eqnarray}
which, since
$\cos(\lambda s^{x})=\cos (\lambda /2)$
and
$\sin (\lambda s^{x})=2\sin (\lambda /2)s^{x}$,
gives
\begin{eqnarray}
\hat{s}_j^z(t)
=
\left(
\cos\frac{tJ_{j-1}}{2}
\cos\frac{tJ_{j}}{2}
-4\sin\frac{tJ_{j-1}}{2}
\sin\frac{tJ_{j}}{2}
s_{j-1}^xs_{j+1}^x
\right)
s_j^z
\nonumber\\
+
2\left(
\sin\frac{tJ_{j-1}}{2}
\cos\frac{tJ_{j}}{2}
s_{j-1}^x
+
\cos\frac{tJ_{j-1}}{2}
\sin\frac{tJ_{j}}{2}
s_{j+1}^x
\right)
s_j^y.
\end{eqnarray}
With the aid of this expression, the integration in Eq. (5)
is readily perform to yield the desired result
\begin{eqnarray}
\chi_{zz}
=\frac{1}{N}
\sum_{n,p}
\left[
\left(
\frac{\sinh\frac{J_{n-1}-J_n}{2kT}}{J_{n-1}-J_n}
+\frac{\sinh\frac{J_{n-1}+J_n}{2kT}}{J_{n-1}+J_n}
\right)
\langle s_n^zs_p^z\rangle_0
\right.
\nonumber\\
\left.
-4
\left(
\frac{\sinh\frac{J_{n-1}-J_n}{2kT}}{J_{n-1}-J_n}
-\frac{\sinh\frac{J_{n-1}+J_n}{2kT}}{J_{n-1}+J_n}
\right)
\langle s_{n-1}^xs_n^zs_{n+1}^xs_p^z\rangle_0
\right.
\nonumber\\
\left.
-2\begin{rm}i\end{rm}
\left(
\frac{\cosh\frac{J_{n-1}-J_n}{2kT}-1}{J_{n-1}-J_n}
+\frac{\cosh\frac{J_{n-1}+J_n}{2kT}-1}{J_{n-1}+J_n}
\right)
\langle s_{n-1}^xs_n^ys_p^z\rangle_0
\right.
\nonumber\\
\left.
+2\begin{rm}i\end{rm}
\left(
\frac{\cosh\frac{J_{n-1}-J_n}{2kT}-1}{J_{n-1}-J_n}
-\frac{\cosh\frac{J_{n-1}+J_n}{2kT}-1}{J_{n-1}+J_n}
\right)
\langle s_{n}^ys_{n+1}^xs_p^z\rangle_0
\right].
\end{eqnarray}

Since $H_{0}$ does not contain $s^{y}$ or $s^{z}$, and
$s^{y}s^{z}=(\begin{rm}i\end{rm}/2)s^{x}$,
Tr$s^{\alpha }=0$ and
$(s^{\alpha })^{2}=1/4$, it is readily seen on using a
representation in which $s^{x}$ is diagonal that the correlation
functions appearing in Eq. (8) can be
simplified. Namely,
$\langle s_n^zs_p^z\rangle_0=(1/4)\delta_{np}$,
$\langle s_{n-1}^xs_n^zs_{n+1}^xs_p^z\rangle_0
=(1/4)\delta_{np}\langle s_{n-1}^xs_{n+1}^x\rangle_0$,
$\langle s_{n-1}^xs_n^ys_p^z\rangle_0
=(\begin{rm}i\end{rm}/2)\delta_{np}
\langle s_{n-1}^xs_n^x\rangle_0$,
$\langle s_{n}^ys_{n+1}^xs_p^z\rangle_0
=(\begin{rm}i\end{rm}/2)\delta_{np}
\langle s_{n}^xs_{n+1}^x\rangle_0$.
The pair correlation functions in $s^{x}$ that remain may then
be evaluted using Eq. (2). In
this way, it follows
after straightforward calculations,
that the initial longitudinal and transverse
static susceptibilities, Eq. (4) and Eq. (8), of
the nonuniform Ising chain
may be expressed in the form
\begin{eqnarray}
\chi_{xx}
=\frac{1}{4kT}
\left[
1+\frac{2}{N}
\sum_n\sum_{p>n}
\left(-\tanh\frac{J_n}{4kT}\right)
\left(-\tanh\frac{J_{n+1}}{4kT}\right)
\ldots
\left(-\tanh\frac{J_{p-1}}{4kT}\right)
\right];
\end{eqnarray}
\begin{equation}
\chi_{zz}
=\frac{1}{2N}\sum_n
\left(
\frac{\tanh\frac{J_{n-1}}{4kT}-\tanh\frac{J_{n}}{4kT}}{J_{n-1}-J_{n}}
+\frac{\tanh\frac{J_{n-1}}{4kT}+\tanh\frac{J_{n}}{4kT}}{J_{n-1}+J_n}
\right).
\end{equation}
The result, Eq. (10), for $\chi_{zz}$
coincides with the result obtained
by Idogaki, Rikitoku and Tucker \cite{irt}
using a different approach.

The off-diagonal components $\chi _{\alpha \beta }$ can be shown
to be zero as follows. $\chi _{\beta \alpha }$ contains
\linebreak
Tr$[\exp(-H_{0}/kT)\hat{s}^{\beta }(t)s^{\alpha }]$.
Again, using a representation in which $s^{x}$
is diagonal, this trace is zero if $\beta =x$ and $\alpha =y$ or $z$.
Thus $\chi _{xy}$ and $\chi _{xz}$ are zero. Also, from the definition
of the interaction representation, $\hat{s}(t)$, and the
fact that a trace is invarient
under cyclic rotation of the operators under it, it follows that
$\chi _{yx}$ and $\chi _{zx}$ are also zero. Further, $\chi _{zy}$
is given by Eq. (8) with $s_{p}^{z}$ replaced by $s_{p}^{y}$. All
the correlation functions on the right hand side then reduce to zero, so
$\chi _{zy}=0$. Finally, $\chi _{yz}=\chi_{zy}$ ($=0$) and $\chi _{yy}
=\chi _{zz}$ from the symmetry transformation
${s^x}^{\prime}=-s^x$,
${s^y}^{\prime}=s^z$,
${s^z}^{\prime}=s^y$.

One immediately observes for the regular chain, $J_1=J_2=\ldots\;$,
that Eq. (9) gives \cite{s,b}
\begin{eqnarray}
\chi_{xx}^{J_1}
=\frac{1}{4kT}
\left[
1+\frac{2}{N}
\sum_{q}(N-q)
\left(-\tanh\frac{J_1}{4kT}\right)^q
\right]
\nonumber\\
=\frac{1}{4kT}
\left(
1-\frac{2\tanh\frac{J_1}{4kT}}{1+\tanh\frac{J_1}{4kT}}
\right)
=\frac{{\mbox {e}}^{-\frac{J_1}{2kT}}}{4kT},
\end{eqnarray}
in the thermodynamic limit.
For the regular alternating-bond chain,
$J_1=J_3=\ldots\;$, $J_2=J_4=\ldots\;$, Eq. (9) yields
\begin{equation}
\chi_{xx}^{J_1J_2}
=\frac{1}{4kT}
\frac{\left(
1-\tanh\frac{J_1}{4kT}\right)
\left( 1-\tanh\frac{J_2}{4kT}\right)}
{1-\tanh\frac{J_1}{4kT}\tanh\frac{J_2}{4kT}},
\end{equation}
and for the case of the regular alternating-bond chain
$J_1=J_2=J_5=J_6\ldots\;$,
$J_3=J_4=J_7=J_8\ldots\;$
\begin{eqnarray}
\chi_{xx}^{J_1J_1J_3J_3}
=\frac{1}{4kT}
\left[
1-
\frac{\tanh\frac{J_1}{4kT}+\tanh\frac{J_3}{4kT}}
{1-\tanh\frac{J_1}{4kT}\tanh\frac{J_3}{4kT}}
\right.
\nonumber\\
\left.
+\frac{\frac{1}{2}
\left(
\tanh\frac{J_1}{4kT}+\tanh\frac{J_3}{4kT}
\right)^2
+2\tanh^2\frac{J_1}{4kT}\tanh^2\frac{J_3}{4kT}}
{1-\tanh^2\frac{J_1}{4kT}\tanh^2\frac{J_3}{4kT}}
\right].
\end{eqnarray}
Further, for the transverse components, Eq. (10) yields the results:
$\chi_{zz}^{J_1}$ \cite{cds},
$\chi_{zz}^{J_1J_2}$ \cite{irt},
$\chi_{zz}^{J_1J_1J_3J_3}
=\frac{1}{4}\chi_{zz}^{J_1}
+\frac{1}{4}\chi_{zz}^{J_3}
+\frac{1}{2}\chi_{zz}^{J_1J_3}$,
$\chi_{zz}^{J_1J_1J_1J_4J_4J_4}
=\frac{1}{3}\chi_{zz}^{J_1}
+\frac{1}{3}\chi_{zz}^{J_4}
+\frac{1}{3}\chi_{zz}^{J_1J_4}$ etc..

It is worthwhile to note, that $\chi_{zz}$, Eq. (10),
does not depend on the sign of the intersite coupling,
whereas $\chi_{xx}$, Eq. (9), does.
Thus, according to Eq. (11)
for the regular chain,
one finds immediately that for ferromagnetic coupling ($J_1<0$)
$\chi_{xx}\to\infty$ as $T\to 0$,
whereas for antiferromagnetic coupling ($J_1>0$)
$\chi_{xx}\to 0$ as $T\to 0$.
In the case of the regular alternating-bond chain
$J_1=J_3=\ldots\;$,
$J_2=J_4=\ldots\;$,
Eq. (12) implies that $\chi_{xx}$ diverges as $T\to 0$ if both $J_{1}$
and $J_{2}$ are ferromagnetic,
whereas if either, or both, of the intersite couplings is
antiferromagnetic, it does not diverge.
From Eq. (13)
for the alternating-bond chain
$J_1=J_2=J_5=J_6=\ldots\;$,
$J_3=J_4=J_7=J_8=\ldots\;$,
one finds that if both, or either, of $J_{1}$ and $J_{3}$ are ferromagnetic
$\chi_{xx}$ diverges as $T\to 0$,
whereas, it does not, if both couplings are antiferromagnetic.
These results are to be expected if one considers the ground state of
these regular alternating-bond spin chains.
For example, it is clear that the ground state
of the chain
$J_1=J_3=\ldots\;$,
$J_2=J_4=\ldots\;$, with
$J_1<0, J_2>0$,
is of an antiferromagnetic type,
whereas the ground state of the chain
$J_1=J_2=J_5=J_6=\ldots\;$,
$J_3=J_4=J_7=J_8=\ldots\;$, having
$J_1<0, J_3>0$,
is ferromagnetic in character.

\section{Initial static susceptibilities
         \protect\\
         of random Ising chains}

Consider an Ising chain in which the exchange couplings
are random variables evenly distributed with a probability
$p(J_1,\ldots,J_{N-1})=\prod_np(J_n)$.
The random-averaged susceptibilities are given by
\begin{eqnarray}
\overline{\chi_{xx}}
=\frac{1}{4kT}
\left[
1+\frac{2}{N}
\sum_{q}(N-q)
\overline{\left(-\tanh\frac{J}{4kT}\right)}^q\;
\right]
\nonumber\\
=\frac{1}{4kT}
\left(
1-\frac{2\overline{\tanh\frac{J}{4kT}}}
{1+\overline{\tanh\frac{J}{4kT}}}
\right)
\end{eqnarray}
(in the first equality
use has been made of the fact that
$\overline{-\tanh(J_p/4kT)}$
is the same for all sites $p$), and
\begin{equation}
\overline{\chi_{zz}}
=\frac{1}{2}
\left[
\overline{
\left(
\frac{\tanh\frac{J_{1}}{4kT}-\tanh\frac{J_{2}}{4kT}}{J_{1}-J_{2}}
\right)}
+
\overline{
\left(
\frac{\tanh\frac{J_{1}}{4kT}+\tanh\frac{J_{2}}{4kT}}{J_{1}+J_{2}}
\right)}
\right] .
\end{equation}
Besides the probability distribution
\begin{eqnarray}
p(J_n)=c\delta(J_n-J_1)+(1-c)\delta(J_n-J_2),
\;\;\;
0\le c\le 1
\end{eqnarray}
for which
$\overline{\chi_{zz}}$
was examined in Ref. \cite{irt},
the Gaussian distribution
\begin{eqnarray}
p(J_n)=\frac{1}{\sqrt{2\pi}\sigma}
\exp \left[ -\frac{(J_{n}-J_{0})^{2}}{2\sigma ^{2}}\right]
\end{eqnarray}
and the Lorentzian distribution
\begin{eqnarray}
p(J_n)=\frac{1}{\pi}
\frac{\Gamma}{(J_n-J_0)^2+\Gamma^2}
\end{eqnarray}
centred at $J_0$ with a strength of disorder
controlled by $\sigma^2$ and $\Gamma$, respectively, will be considered.

Consider first the longitudinal susceptibility
$\overline{\chi_{xx}}$, Eq. (14), (Figs. 1-3).
The low-temperature behaviour, discussed in Section 2,
for the uniform chain arose because $\tanh(J/4kT)\to {\mbox{sgn}}J$ as
$T\to 0$.
As a result, in this limit $\chi_{xx}$ diverges
more rapidly than $1/T$ for $J<0$, but does not diverge for $J>0$.
For the random case, the essential quantity influencing the initial
longitudinal static susceptibility
is $\mid \overline{\tanh (J/4kT)}\mid $.
If there are present, exchange interactions of opposite sign,
then this quantity is  $<1$ as $T\to 0$
and a qualitative change in the temperature dependence of $\chi _{xx}$
from that of the uniform chain results.
The low temperature behaviour is now of the paramagnetic type,
$1/T$, (the dotted lines in Figs. 1b,
1c and 2). A similar qualitative change
in the temperature dependence occurs
in the presence of randomness defined by Eqs. (17) or (18)
(the dotted lines in Fig. 3). On the other hand, in the presence of
any randomness defined by Eq. (16)
in which $J_{1}$ and $J_{2}$ have the same
sign, the quantity $\mid \overline{\tanh (J/4kT)}\mid$ still equals
unity in the limit $T\to 0$, and thus this randomness only leads to
a quantitative change in the thermal dependence of
$\overline{\chi_{xx}}$ from that of the uniform case (the dotted lines in
Figs. 1a, 1d).
In Figs. 2 and 3 one can also trace a crossover
from a ferromagnetic or antiferromagnetic type
of thermal behaviour
at high temperatures to a paramagnetic type
of thermal behaviour
as $T\to 0$, which occurs with the introduction of random-bond disorder
given by distributions of Eqs.
(16) with $J_1J_2<0$,
(17) or (18).
Any small amount of disorder causes this crossover,
but the smaller the disorder
the lower is the temperatures at which it occurs.

It is also of interest to compare the results for the alternating-bond
chains of Section 2 with that for the random chain (described by Eq. (16))
having equal concentrations of the two types of bond (i.e. $c=0.5$).
It was seen for the regular alternating-bond models
having $J_{1}>0$ and $J_{2}<0$
(or vice versa), that the low temperature longitudinal susceptibility either
did (in the case of $J_{1}J_{1}J_{2}J_{2}\ldots $), or did not
(in the case of $J_{1}J_{2}J_{1}J_{2}\ldots $) diverge at low temperatures.
Further, the divergence when it occurs, is
greater than $1/T$.
For the corresponding random chain ($c=0.5$) having these
signs for the exchange interactions,
it has been observed above that a $1/T$
dependence results instead. (Compare the short-dashed and long-dashed
lines with the curve for $c=0.5$ in Figs. 1b, 1c and 2). On the other hand,
when $J_{1}$ and $J_{2}$ are either both positive or negative, the
corresponding curves only differ quantitatively (Figs. 1a and 1d).

Finally, it is noted that the initial longitudinal static susceptibility
for chains with Gaussian and Lorentzian disorders having $J_{0}=0$
exhibit the temperature dependence of an
ideal paramagnetic
$\chi_{xx}=1/4kT$, independent of the strength of the disorder.

Consider now the thermal behaviour of the initial transverse static
susceptibility $\overline{\chi_{zz}}$, Eq. (15), (Figs. 4-7).
For the uniform Ising chain it is finite at all temperatures; equal
to $1/(2J)$ at $T=0$, slightly increasing with increase
of temperature, and then decreasing to approach the temperature
dependence characteristic of an ideal paramagnet at high temperatures.
The weaker the exchange interaction between the longitudinal
($x$) components of the spins,
the larger is the value of $\chi _{zz}$ at zero
temperature, the smaller is the temperature at which $\chi _{zz}$ exhibits
its maximum, and the smaller is the temperature at which it exhibits
behaviour characteristic of an ideal paramagnet. These features will be
reflected in the sequence of curves for the temperature dependence
of $\overline{\chi _{zz}}$ of the random chain, described by
Eq. (16), as the concentration is varied. Fig. 4 shows such a series of
curves as $c$ varies from 1 to 0 for the particular case when
$J_{1}=\mid J\mid$
and
$J_{2}=0.3\mid J\mid$.
The susceptibilities of some regular alternating-bond
chains are also depicted. It is interesting to note that the initial
transverse static susceptibility for the random-bond chain having $c=0.5$ is
enhanced over that for the alternating-bond chain
$J_{1}J_{2}J_{1}J_{2}\ldots \;$,
coincides with that for the chain
$J_{1}J_{1}J_{2}J_{2}J_{1}J_{1}J_{2}J_{2}\ldots \;$,
and is suppressed in comparison to that for the chain
$J_{1}J_{1}J_{1}J_{2}J_{2}J_{2}\ldots \;$,
etc..

Figs. 5,6 illustrate a
difference in the influence of Gaussian, Eq. (17), and Lorentzian, Eq. (18),
bond disorder on the temperature dependence of
$\overline{\chi_{zz}}$.
For increasing, but small, Gaussian disorder strength an increasing
enhancement of $\overline{\chi_{zz}}$ occurs at low temperatures
(curves 2-4 in Fig. 5). Above a certain disorder strength,
this low temperature
enhancement is reduced (curves 5,6), and finally a suppression of
$\overline{\chi_{zz}}$ below its non-disordered value occurs (curve 7).
In contrast to this behaviour for Gaussian disorder,
chains with Lorentzian disorder exhibit a decrease
in $\overline{\chi_{zz}}$ with
increasing disorder strength for almost all temperatures (Fig. 6).

Finally, it is noted that in contrast to the initial longitudinal
static susceptibility, the initial transverse static susceptibility for
chains having  Gaussian or Lorentzian disorder, with $J_{0}=0$,
decreases with increase of
disorder strength (Fig. 7). This is not surprising, since
the larger the exchange coupling, the smaller is $\chi _{zz}$,
and with increasing
$\sigma ^{2}$ or $\Gamma $ larger values of exchange couplings are
present more often.

\section{Discussion}

To summarise.
An alternative, but more natural derivation than that of Ref. \cite{irt},
of the exact formulae for the initial static susceptibility
of random-bond Ising chains has been presented.
The former requires evaluation of the thermal averaged magnetization
to first-order in the magnetic field strength,
whereas the latter, requires evaluation of just the
field-independent part of the double-time pair correlation function.
The latter approach also permits one to derive similarly
the frequency dependent susceptibility and the structure factor
of nonuniform and random spin-$\frac{1}{2}$ Ising chains;
work in this direction is in progress.
Dependent on the type of randomness present,
it has been demonstrated
that the disorder may lead to either a qualitative, or
to only  a quantitative change,
in the thermal dependence of the initial static susceptibilities.
Even a small amount of disorder may lead to a qualitative
deviation from that of the nonrandom case
at low temperatures.
A difference in the temperature dependence
of the initial static susceptibilities of
regular alternating-bond chains
having two types of exchange bonds,
and the corresponding
random-bond chain, has been revealed.
Comparison of the numerical results for chains having
Gaussian bond-disorder and for those having Lorentzian bond-disorder
show a marked difference in the dependence of the
transverse susceptibility on the strength of the disorder.
Although, the theoretical results observed in our work
should prove valuable in understanding the effects of disorder on the
observable properties of Ising chain materials,
there are to our knowledge, no experimental results
yet available
that enable a direct comparison between theory and experiment to be made.
However, with for example,
the synthesis of magnetic chain molecular materials becoming a reality,
this lack of experimental data may be resolved in the future.

\vspace{5mm}

A part of this work was done during the Summer College in Condensed Matter
on ``Statistical Physics of Frustrated Systems'' (July - August 1997) at
the ICTP, Trieste. One of the authors (O.D.) is grateful to the ICTP for
support and hospitality.

\clearpage

\noindent
{\bf List of figure captions}

\vspace{0.7cm}

\noindent
Fig. 1.  Thermal dependence
of the initial longitudinal static susceptibility of
disordered-bond (described by Eq. (16))
and regular alternating-bond Ising chains when
$J_{1}$ and $J_{2}$ are respectively,
(a) $-\mid J\mid$, $-0.3\mid J\mid$,
(b) $-\mid J\mid$, $ 0.3\mid J\mid$,
(c) $ \mid J\mid$, $-0.3\mid J\mid$
and
(d) $ \mid J\mid$, $ 0.3\mid J\mid$.
Curve 1: uniform chain, $c=1$. Curves 2 and 3: regular alternating chains
$J_{1}J_{2}J_{1}J_{2}\ldots \;$,
$J_{1}J_{1}J_{2}J_{2}\ldots \;$,
respectively.
Curve 4: disordered chain, $c=0.5$.
Curve 5: uniform chain, $c=0$.
Curve 6: ideal paramagnet, $J_{1}=J_{2}=0$.

\vspace{0.7cm}

\noindent
Fig. 2. Thermal dependence of
$kT\overline{\chi _{xx}}$ for disordered-bond
(described by Eq. (16)) and regular alternating-bond Ising chains when
$J_{1}=-\mid J\mid$
and
$J_{2}=0.3\mid J\mid$.
Curve 1: uniform chain, $c=1$.
Curves 2-8:
disordered chains,
$c=$ $0.9$, $0.8$, $0.7$, $0.5$, $0.3$, $0.2$ and $0.1$,
respectively.
Curve 9: uniform chain $c=0$.
Curves 10 and 11:
regular alternating-bond chains
$J_{1}J_{2}J_{1}J_{2}\ldots $
and
$J_{1}J_{1}J_{2}J_{2}\ldots \;$,
respectively.
Curve 12: ideal paramagnet,
$J_{1}=J_{2}=0$.

\vspace{0.7cm}

\noindent
Fig. 3.  Thermal dependence of
$kT\overline{\chi_{xx}}$ for Ising chains
with (a) Gaussian
and (b) Lorentzian bond disorder.
Curves 1 to 3 are for
$J_{0}=-\mid J\mid$
with
$\left( \sigma /\mid J\mid\right)^{2}=0$
or
$\Gamma /\mid J\mid =0$,
$\left( \sigma /\mid J\mid\right)^{2}=0.5$
or
$\Gamma /\mid J\mid =0.5$
and
$\left( \sigma /\mid J\mid\right)^{2}=1$
or
$\Gamma /\mid J\mid =1$,
as appropriate to (a) and (b), respectively.
Curves 4 to 6 are for
$J_{0}=\mid J\mid$,
with the same pairs of
$\left( \sigma /\mid J\mid\right)^{2}$
and $\Gamma /\mid J\mid$
as for curves 1 to 3, respectively.
Curve 7 is for the ideal paramagnet,
$J_{0}=0$.

\vspace{0.7cm}

\noindent
Fig. 4.  Thermal dependence  of the initial transverse static
susceptibility
of disordered-bond (described by Eq. (16))
and regular alternating-bond
Ising chains
when
$J_{1}=\mid J\mid$
and
$J_{2}=0.3\mid J\mid$.
Curve 1: uniform chain, $c=1$.
Curves 2 to 4: disordered chains, $c=$ $0.7$, $0.5$, $0.3$, respectively.
Curve 5: uniform chain $c=0$.
Curves 6 to 8: regular alternating-bond chains
$J_{1}J_{2}J_{1}J_{2}\ldots \;$,
$J_{1}J_{1}J_{2}J_{2}\ldots $
and
$J_{1}J_{1}J_{1}J_{2}J_{2}J_{2}\ldots \;$,
respectively.

\vspace{0.7cm}

\noindent
Fig. 5.  Thermal dependence of the initial transverse static
susceptibility
of Ising chains with Gaussian bond-disorder, when
$J_{0}=\mid J\mid$.
Curve 1: no disorder,
$\left( \sigma /\mid J\mid\right)^{2}=0$.
Curves 2 to 7:
$\left( \sigma /\mid J\mid\right)^{2}=$
$0.1$, $0.2$, $0.5$, $1$, $2$ and $4$, respectively.

\vspace{0.7cm}

\noindent
Fig. 6.  Thermal dependence of the initial transverse susceptibility
of Ising chains with Lorentzian bond-disorder, when
$J_{0}=\mid J\mid$.
Curve 1: no disorder,
$\Gamma /\mid J\mid =0$.
Curves 2-6,
$\Gamma /\mid J\mid =$
$0.02$, $0.1$, $0.2$, $0.5$ and $1$, respectively.

\vspace{0.7cm}

\noindent
Fig. 7.  Thermal dependence of the initial transverse susceptibility
of Ising chains with
(a) Gaussian
and (b) Lorentzian bond-disorder,
when
$J_{0}=0$.
Curve 1: ideal paramagnet,
$\left( \sigma /\mid J\mid\right)^{2}=0$
or
$\Gamma /\mid J\mid =0$,
as appropriate in (a) and (b).
Curves 2 and 3:
$\left( \sigma /\mid J\mid\right)^{2}=0.5$
or
$\Gamma /\mid J\mid =0.5$
and
$\left( \sigma /\mid J\mid\right)^{2}=1$
or
$\Gamma /\mid J\mid =1$,
respectively.

\end{document}